\documentclass[amt,noline]{copernicus}

\hack{\usepackage{empheq}}

\begin{document}

\title{Functional derivatives applied to error propagation of uncertainties in topography to large-aperture scintillometer-derived heat fluxes}

\Author[1]{M.~A.}{Gruber}
\Author[1]{G.~J.}{Fochesatto}
\Author[2]{O.~K.}{Hartogensis}
\Author[3]{M.}{Lysy}

\affil[1]{Department of Atmospheric Sciences, College of Natural Science and Mathematics, Geophysical Institute, \hack{\newline} University of Alaska Fairbanks, Fairbanks, USA}
\affil[2]{Meteorology and Air Quality Group, Wageningen University, Wageningen, the Netherlands}
\affil[3]{Department of Statistics and Actuarial Science, University of Waterloo, Waterloo, Canada}

\runningtitle{Error propagation of uncertainties in topography to
scintillometer-derived heat fluxes}

\runningauthor{M.~A.~Gruber et~al.}

\correspondence{M.~A.~Gruber (matthewgruber@gi.alaska.edu)}

\received{12 November 2013}
\pubdiscuss{7 January 2014}
\revised{18 June 2014}
\accepted{21 June 2014 }

\published{}

\firstpage{1}

\maketitle

\hack{\allowdisplaybreaks}
\begin{abstract}
 Scintillometer measurements
allow for estimations of the refractive index structure parameter $C_n^2$
over large areas in the atmospheric surface layer. Turbulent fluxes of heat
and momentum are inferred through coupled sets of equations derived from the
Monin--Obukhov similarity hypothesis. One-dimensional sensitivity functions
have been produced that relate the sensitivity of heat fluxes to
uncertainties in single values of beam height over flat terrain. However,
real field sites include variable topography. We develop here, using
functional derivatives, the first analysis of the sensitivity of
scintillometer-derived sensible heat fluxes to uncertainties in spatially
distributed topographic measurements. Sensitivity is shown to be concentrated
in areas near the center of the beam path and where the underlying topography
is closest to the beam height. Relative uncertainty contributions to the
sensible heat flux from uncertainties in topography can reach 20\,\% of the
heat flux in some cases. Uncertainty may be greatly reduced by focusing
accurate topographic measurements in these specific areas. A~new
two-dimensional variable terrain sensitivity function is developed for
quantitative error analysis. This function is compared with the previous
one-dimensional sensitivity function for the same measurement strategy over
flat terrain. Additionally, a~new method of solution to the set of coupled
equations is produced that eliminates computational error.
\end{abstract}

\hack{\newpage}

\introduction

Large-aperture scintillometers infer the index of refraction structure
parameter $C_n^2$ over large areas of terrain in the atmospheric surface
layer. The structure parameter for temperature $C_{T}^2$ is resolved, and
this information solves for the sensible heat flux $H_\mathrm{S}$ through the
application of equations derived from the Monin--Obukhov similarity
hypothesis \citep{HARTOGENSIS2003, MOENE2003}. The sensible heat flux in the
\mbox{atmospheric} surface layer is given by
\begin{align}
&H_\mathrm{S}=-\rho c_{p} u_\star T_\star, \label{SENSIBLEHEATch3}
\end{align}
where $\rho$ is the density of air, $c_{p}$ is the heat capacity at constant
pressure, $u_\star$ is the friction velocity, and $T_\star$ is the
temperature scale \citep[e.g.,][]{MO1954,OBUKHOV1971,SORBJAN,50YRSMO}. The
temperature scale $T_\star$ is resolved by
\begin{empheq}[left={T_\star = \empheqlbrace}]{alignat=2}
%\[ %
%
& \pm \sqrt{\frac{C_{T}^2} {a}}{z_{\text{eff}}}^{1/3}(1-b\zeta)^{1/3} & \quad\quad\quad \zeta\leq0 , \label{unstabletstar} \\
& \pm \sqrt{\frac{C_{T}^2}{a}}\frac{{z_{\text{eff}}}^{1/3}}{(1+c\zeta^{2/3})^{1/2}} & \quad\quad\quad \zeta\geq0  ,\label{stabletstar}
%\]
\end{empheq}
where $z_{\text{eff}}$ is the effective beam height above the
ground, $\zeta\equiv z_{\text{eff}}/l$, where $l$ is the Obukhov length
\citep[e.g.,][]{SORBJAN}, and $a$, $b$ and $c$ are empirical parameters. The
values of the empirical parameters are taken to be $a=4.9$, $b=6.1$, and
$c=2.2$, as seen in \citet{ANDREAS1989} after an adjustment from the original
values seen in \cite{WYNGARD1971}. These values may not be appropriate for
all field sites. We will assume that $C_{T}^2$ is resolved by neglecting the
influence of humidity fluctuations, although this does not influence our
results.

As can be imagined from Eqs.~(\ref{unstabletstar}) and (\ref{stabletstar}),
it is important to know the height $z$ at which $C_{T}^2$ is being sampled;
this corresponds to the scintillometer beam height. The beam height usually
varies along the beam path. Even if turbulence is being sampled above an
extremely flat field, uncertainty in $z$ will still be present. Previous
studies such as \citet{ANDREAS1989} and \citet{HARTOGENSIS2003} have
quantified the sensitivity of $H_\mathrm{S}$ to uncertainties in $z$ over
flat terrain. It is the goal of this study to extend the theoretical
uncertainty analysis of \citet{ANDREAS1989} and \citet{HARTOGENSIS2003} to
take into account variable terrain along the path. The value of this is in
the ability to evaluate uncertainty estimates for scintillometer measurements
over variable terrain, as well as to study the theoretical effect that the
underlying terrain has on this uncertainty.

The studies of \citet{ANDREAS1989} and \citet{HARTOGENSIS2003} assume an
independently measured friction velocity $u_\star$. With large-aperture
scintillometers, $u_\star$ may be inferred through the Businger--Dyer
relation of wind stress, which is coupled to the Monin--Obukhov equations
\citep[e.g.,][]{HARTOGENSIS2003,SOLIGNAC}. Alternatively, with displaced-beam
scintillometers, path-averaged values of the inner-scale length of turbulence
$l_{o}$ can be measured (in addition to $C_n^2$), which are related to the
turbulent dissipation rate $\epsilon$, which is in turn related through
coupled Monin--Obukhov equations to $u_\star$ \citep[e.g.,][]{ANDREAS1992}.
As a~first step towards a~variable terrain sensitivity analysis for
large-aperture scintillometers, we will assume independent $u_\star$
measurements such that the Businger--Dyer equation will not be considered.
Additionally, in order to take into account thick vegetation, the
displacement distance $d$ is often introduced. We will not consider this for
the purposes of this study.

We are thus considering a~large-aperture scintillometer strategy with
independent $u_\star$ measurements as in \citet{ANDREAS1989} and Appendix
A~of \citet{HARTOGENSIS2003}, and we consider the line integral effective
beam height formulation from \citet{HARTOGENSIS2003} and \citet{KLEISSL2008}.
The effective height formulation is also discussed in \citet{EVANS2011} and
in \citet{GELI2012}. The assumptions behind this line integral approach are
that the profile of $C_{T}^2$ above the ground satisfies the Monin--Obukhov
profile at any point along the beam path, and also that $H_\mathrm{S}$ is
constant vertically and horizontally within the surface layer region sampled
by the beam. In this case, two coupled effects must be taken into account.
Firstly, the scintillometer is most sensitive to fluctuations in the index of
refraction towards the center of its beam. This is due to the optical
configuration of the scintillometer system; a~unit-less optical path
weighting function takes this into account
\citep[e.g.,][]{OCHSPWF,HARTOGENSIS2003}. The second effect is that, in areas
where the topography approaches the beam, the $C_{T}^2$ being sampled is
theoretically more intense than in areas where the terrain dips farther below
the beam.

In Sect.~2 of this paper, we define the sensitivity function
$S_{H_\mathrm{S},z}(u)$ for the sensible heat flux $H_\mathrm{S}$ as
a~function of variable topography $z(u)$, where $u$ is the relative path
position along the beam. In Sect.~3, we solve for $S_{H_\mathrm{S},z}(u)$ for
any general given $z(u)$. In Sect.~4 we visualize the results by applying the
resulting sensitivity function to the topography of a~real field site in the
North Slope of Alaska. We then apply the resulting sensitivity function to
examples of synthetic beam paths. In Sect.~5 we discuss our results, and we
conclude in Sect.~6.

\section{Definition of the sensitivity function $S_{H_\mathrm{S},z}(u)$}

Under stable conditions ($\zeta>0$), the set of equations to consider
consists of Eqs.~(\ref{SENSIBLEHEATch3}) and (\ref{stabletstar}), as well
as
\begin{align}
&
\zeta= \frac{\kappa g T_\star z_{\text{eff}}}{{u_\star}^2T} \label{zeta},\\
&z_{\text{eff}}= \left(\int_0^1{{z(u)}^{-2/3}G(u)\mathrm{d}u}\right)^{-3/2} \label{stablezeff},
\end{align}
where $z_{\text{eff}}$ is derived in \citet{KLEISSL2008} based on the theory
from \citet{HARTOGENSIS2003}, $z(u)$ is the height of the beam along the
relative path position $u$, $T$ is the temperature, $G(u)$ is the optical
path weighting function, $g$ is gravitational acceleration, and $\kappa$ is
the von K\'arm\'an constant ($0.4$).

For unstable conditions ($\zeta<0$), Eqs.~(\ref{SENSIBLEHEATch3}),
(\ref{unstabletstar}) and (\ref{zeta}) are considered, but
Eq.~(\ref{stablezeff}) is replaced by
\begin{align}
&z_{\text{eff}}=
\nonumber\\&\hack{\hbox\bgroup\fontsize{7.5}{7.5}\selectfont$\displaystyle}\frac{z_{\text{eff}}}{2b\zeta}\left(1-\sqrt{1-\frac{4b\zeta}{z_{\text{eff}}}\left[\int_0^1{{z(u)}^{-2/3}\left(1-b\zeta\frac{z(u)}{z_{\text{eff}}}\right)^{-2/3}G(u)\mathrm{d}u}\right]^{-3/2}}\right),
\label{zeffunstable} \hack{$\egroup}\end{align} where $z_{\text{eff}}$ is
derived in \citet{HARTOGENSIS2003}.

The propagation of uncertainty from measurements such as $z(u)$ to derived
variables such as $H_\mathrm{S}$ will be evaluated in the context of the
inherent assumptions behind the theoretical equations. A standard
approximation~(e.g., Taylor, 1997) to
the uncertainty in estimating the derived variable $f = f(\mu)$, $\mu =
(\mu_1, \mu_2, \ldots, \mu_N)$, by $\hat f = f(x)$, a function of measurement
variables $x = (x_1, x_2, \ldots, x_N)$, is
%\correct{A common} estimate of the uncertainty in the derived variables that
%results from small errors in measurements \correct{\citep[e.g.][]{TAYLOR}} is given by
\begin{align}\label{errorpropch3}
{\sigma^2_f} = &E\big\{[f(x) - f(\mu)]^2\big\} \approx \sum_{i=1}^N\left({\frac{\partial f}{\partial x_i}}\right)^2E\big[(x_i-\mu_i^2)\big] \nonumber\\&= \sum_{i=1}^N\left({\frac{\partial f}{\partial x_i}}\right)^2\sigma_i^2.
\end{align}
%\correct{Here, the derived variable $f = f(x)$ is a function of measurement variables $x = (x_1,x_2,\dots, x_N)$, each having independent Normally distributed uncertainties, $x_i \stackrel{\textrm{iid}}{\sim} \mathcal N(\mu_i, \sigma_i^2)$}.

The numerical indices indicate different independent (measurement) variables,
such as $T$, $P$, $C_n^2$, $u_\star$, and beam wavelengths $\lambda$ and $z$.
%The first and last terms in Eq.~(\ref{errorpropch3}) represent an offset from the true solution (inaccuracy), whereas the central square root term represents the breadth of uncertainty due to random error (imprecision).
It is convenient to re-write Eq.~(\ref{errorpropch3}) as
\begin{align}
 &
 \left(\frac{{\sigma_f}}{f}\right)^2 = \sum_{i=1}^N S_{{f, x_i}}^2\frac{\sigma_i^2}{x_i^2},  \label{errorprop2ch3}
\end{align}
where the sensitivity functions $S_{f, x} =$ $(S_{{f, x_1}}, S_{{f, x_2}},
\ldots, S_{{f, x_N}})$ are defined as
\begin{align}
 &
S_{{f, x_i}} \equiv \frac{x_i}{f}\left(\frac{\partial f}{\partial
x_i}\right). \label{SENSITIVITYEQch3}
\end{align}

Sensitivity functions such as these are developed in \citet{ANDREAS1989} and
\citet{ANDREAS1992}. They are each a~measure of the portion of relative error
in a~derived variable $f$ resulting from a relative error in the individual
measurement variable $x_i$. The problem of resolving the uncertainty in the
derived variables is a~matter of identifying the magnitude and character of
the measurement uncertainties, and then solving for the partial derivative
terms in Eqs.~(\ref{errorpropch3}) and (\ref{SENSITIVITYEQch3}).

Here we seek a~solution to the sensitivity function of sensible heat flux as
a~function of topography $S_{H_\mathrm{S},z}$. In the flat terrain case, the
sensitivity function $S_{H_\mathrm{S},z}$ has a single component,
corresponding to the single measurement variable $z$~\citep{ANDREAS1989}. In
our situation, however, we may imagine that since $z(u)$ is distributed over
one dimension instead of a~single value of $z$, $S_{H_\mathrm{S},z}$ will be
composed of a spectrum of components:
\begin{align}
S_{H_\mathrm{S}, z} = \{S_{H_\mathrm{S}, z}(u), \ 0 \le u \le 1\}.
\end{align}

We are thus aiming to expand the sensitivity function denoted ``$S_z$'' in
Fig.~4 of \citet{ANDREAS1989} (our $S_{H_\mathrm{S},z}$ in
Fig.~\ref{STZCONSTRATIO}) from one dimension to infinitely many, owing to the
fact that some derived variables such as $z_{\text{eff}}$ are functions of an
integral over continuous variables $z(u)$ and $G(u)$ (we consider for
generality that $z(u)$ has a continuous uncertainty $\sigma(u)^2$). In other
words, $z_{\textrm{eff}} = z_{\textrm{eff}}[z]$ is a
functional,
having argument $z = \{z(u), \ 0 \le u \le 1\}$.

Being dependent on a continuum of measurement variables, the sensitivity
function $S_{H_\mathrm{S}, z}(u)$ here requires the calculation of a
so-called \emph{functional} derivative, $\delta z_\textrm{eff}/\delta
z(u)$~(e.g., Courant, 1953; Greiner and Reinhardt, 1996). Functional
derivatives have a long history of application to statistical error
analysis~(e.g., Fernholz, 1983; Beutner, 2010, and many references therein).

For our purposes, a heuristic derivation of $\delta z_\textrm{eff}/\delta
z(u)$ results from an interpretation of the integral in $z_\textrm{eff}$ as
the limit of Riemann sums. That is,
\begin{align}
z_{\text{eff}} =& \left(\int\limits_0^1{{z(u)}^{-2/3}G(u)\mathrm{d}u}\right)^{-3/2} \equiv\nonumber\\& \left(\lim_{N \to \infty}\sum\limits_{i=1}^N{z_i}^{-2/3}G_i\cdot(1/N)\right)^{-3/2}, \label{discrete}
\end{align}
where subscript $i$ indicates that $u=(i/N)$. Upon discretizing the input
variables, we have
\begin{align}
\left(\frac{\partial z_{\text{eff}}}{\partial z_k}\right) &= \nonumber\\&-\frac{3}{2}
\left(\sum\limits_{i=1}^N{z_i}^{-2/3}G_i\cdot(1/N)\right)^{-5/2}\nonumber\\& \frac{\partial}{\partial z_k}\left(\sum\limits_{i=1}^N{z_i}^{-2/3}G_i\cdot(1/N)\right) \nonumber \\
&= -\frac{3}{2}
\left(\sum\limits_{i=1}^N{z_i}^{-2/3}G_i\cdot(1/N)\right)^{-5/2} \nonumber\\&\quad \times -\frac 2 3 \left({z_k}^{-5/3}G_k\cdot(1/N) \right)\nonumber \\
&= \left(\sum\limits_{i=1}^N{z_i}^{-2/3}G_i\cdot(1/N)\right)^{-5/2}\nonumber\\&{z_k}^{-5/3}G_k\cdot(1/N).
\end{align}

Letting $k = \arg\,\min_k |z(u) - z_k|$ and taking the limit $N \to \infty$,
the desired functional derivative is given by
\begin{align}
\left(\frac{\delta z_{\text{eff}}}{\delta z(u)}\right) & =
  \left(\int\limits_0^1{{z(u)}^{-2/3}G(u)\mathrm{d}u}\right)^{-5/2}{z(u)}^{-5/3}G(u). \label{diracleibnizstable}
\end{align}

We thus define
\begin{align}
 &
S_{H_\mathrm{S}, z}(u)\equiv\frac{z(u)}{H_\mathrm{S}[z]}\left(\frac{\delta H_\mathrm{S}}{\delta z(u)}\right) \label{SHZ}
\end{align}
as the sensitivity function of sensible heat flux $H_\mathrm{S}$ to
uncertainties in variable topography $z(u)$. It is our goal to evaluate
Eq.~(\ref{SHZ}).

\section{Solution of the sensitivity function $S_{H_\mathrm{S},z}(u)$}

\subsection{Stable conditions ($\zeta>0$)}

Under stable conditions, the set of Eqs.~(\ref{SENSIBLEHEATch3}),
(\ref{stabletstar}), (\ref{zeta}) and (\ref{stablezeff}) is coupled in $l$
through $\zeta$; we begin de-coupling them by combining
Eqs.~(\ref{stabletstar}) and (\ref{zeta}) to obtain
\begin{align}
 &
\zeta=(\pm)\frac{\kappa g {z_{\text{eff}}}^{4/3}\sqrt{{C^2_{T}}}}{{u_\star}^2T\sqrt{a}(1+c\zeta^{2/3})^{1/2}}. \label{stablesol}
\end{align}

Since $\zeta>0$, the unsolved sign is positive. With the substitution
\begin{align}
 &
\hat{\Lambda}\equiv\frac{\kappa^2g^2{C^2_{T}}}{{u_\star}^4T^2a},
\end{align}
we re-arrange  Eq.~(\ref{stablesol}) to obtain
\begin{align}
 &
\zeta^2+c\zeta^{8/3}-{\hat{\Lambda}}{z_{\text{eff}}}^{8/3} = 0,\label{zetastableeq}
\end{align}
where $z_{\text{eff}}$ in the stable case is determined by a
priori known functions $z(u)$ and $G(u)$ through
Eq.~(\ref{stablezeff}). The value of $\hat{\Lambda}$, including $C_{T}^2$, is
directly determined from the measurements. The solution of
Eq.~(\ref{zetastableeq}) follows by re-writing it as a~fourth-degree
algebraic equation in $\zeta^{2/3}$:
\begin{align}
 &
(\zeta^{2/3})^3+c(\zeta^{2/3})^4-{\hat{\Lambda}}{z_{\text{eff}}}^{8/3} = 0,\label{zetastableeqalgebraic}
\end{align}
or more practically, it can be solved through fixed-point recursion on the
function
\begin{align}
 &
\zeta=\sqrt{\frac{{\hat{\Lambda}} z_{\text{eff}}^{8/3}}{1+c\zeta^{2/3}}}\equiv \hat{F}(\zeta),        \label{stablerecursion}
\end{align}
where we must consider the positive root. Note that since
Eq.~(\ref{zetastableeqalgebraic}) is fourth degree, Galois theory states that
it has an explicit solution form \citep[e.g.,][]{GALOIS}. It is thus possible
in theory to write $H_\mathrm{S}=h(z(u),C_n^2,P,T,\lambda, u_\star)$, where
$h$ is an explicit function of the measurements; however, it would be quite
an unwieldy equation.

\begin{figure}[t]
\includegraphics[width=85mm]{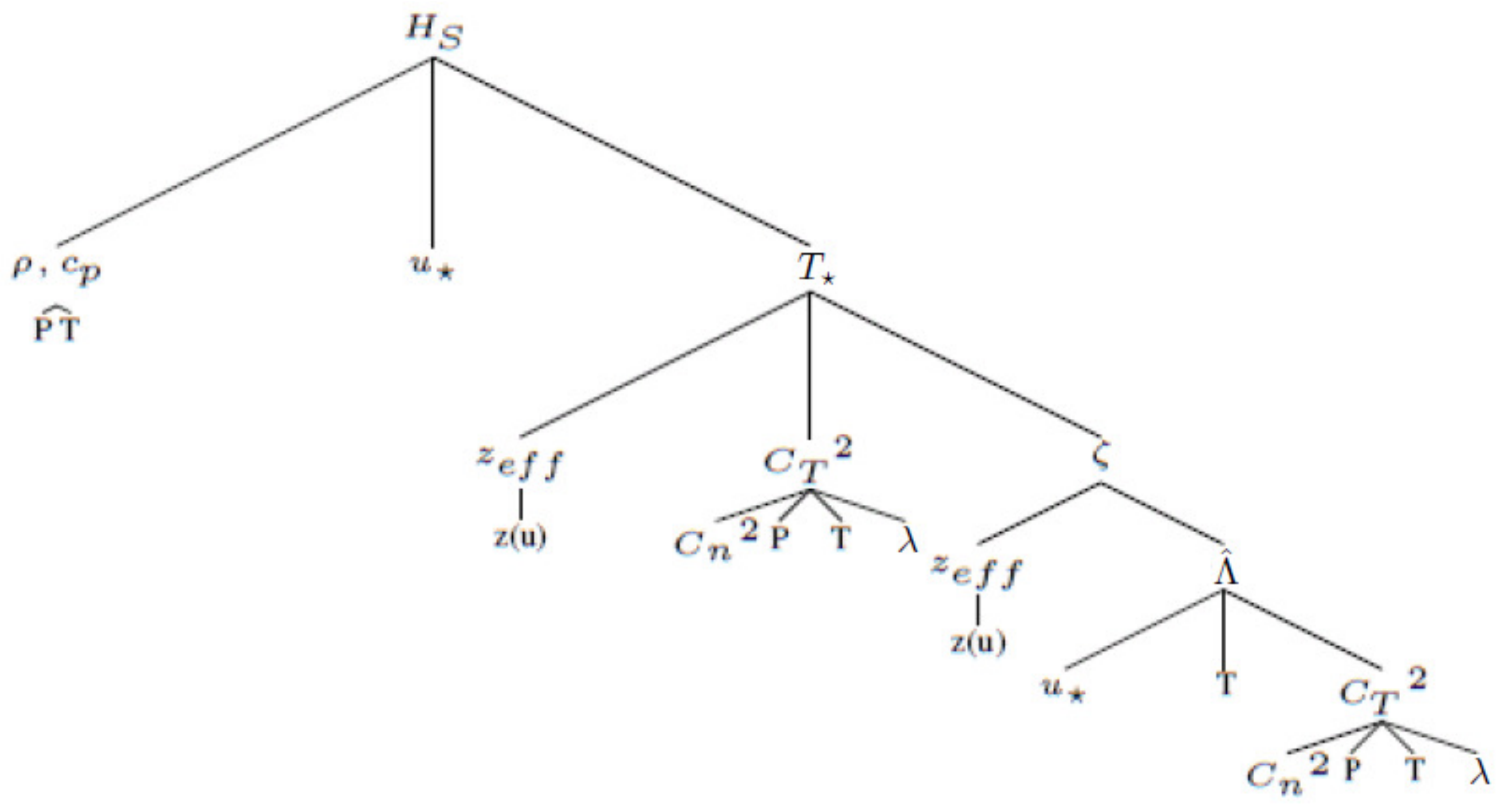}
\caption{Variable inter-dependency tree diagram for the stable case
($\zeta>0$). The measurement variables are at the end of each branch; all
other variables are derived.} \label{TREESTABLE}
\end{figure}

We do not need an explicit solution in order to study the sensitivity; we can
use the chain rule and implicit differentiation as in \citet{GRUBER}. We
establish the variable inter-dependency using Eq.~(\ref{zetastableeq}) as
a~starting point. The tree diagram for any set of measurements under stable
conditions is seen in Fig.~\ref{TREESTABLE}. The measurements are at the ends
of each branch, and all other variables are dependent.

The required global partial derivatives are now defined through the variable
definitions, the above equations, and the tree diagram. We have
\begin{align}
\left(\frac{\delta H_\mathrm{S}}{\delta z(u)}\right) =& \left(\frac{\partial H_\mathrm{S}}{\partial T_\star}\right)\left(\left(\frac{\partial T_\star}{\partial z_{\text{eff}}}\right)_\zeta \right.\nonumber\\&\left.+
\left(\frac{\partial T_\star}{\partial \zeta}\right) \left(\frac{\partial \zeta}{\partial z_{\text{eff}}}\right) \right) \left(\frac{\delta z_{\text{eff}}}{\delta z(u)}\right).\label{exampleerrorzeta}
\end{align}

We will need one derivative that we are not able to retrieve
directly  from explicit definitions. By
implicitly differentiating Eq.~(\ref{zetastableeq}) under the guidance of the
tree diagram seen in Fig.~\ref{TREESTABLE}, we derive
\begin{align}
 &
\left(\frac{\partial \zeta}{\partial z_{\text{eff}}}\right) =\left(\frac{4{\hat{\Lambda}}{z_{\text{eff}}}^{5/3}}{3\zeta+4c\zeta^{5/3}}\right) = \frac{1}{z_{\text{eff}}}\left(\frac{4\zeta(1+c\zeta^{2/3})}{3+4c\zeta^{2/3}}\right).
\end{align}
The functional derivative term $\left(\frac{\delta z_{\text{eff}}}{\delta
z(u)}\right)$ for stable conditions has been evaluated in
Eq.~(\ref{diracleibnizstable}).

\subsection{Unstable conditions ($\zeta<0$)}

Under unstable conditions, the set of Eqs.~(\ref{SENSIBLEHEATch3}),
(\ref{unstabletstar}), (\ref{zeta}) and (\ref{zeffunstable}) is coupled in
$l$ through $\zeta$; note that $z_{\text{eff}}$ is coupled to $\zeta$ in the
unstable case. We combine Eqs.~(\ref{unstabletstar}) and (\ref{zeta}) to
obtain
\begin{align}
 &
\zeta=(\pm)\frac{\kappa g\sqrt{{C^2_{T}}}}{{u_\star}^2T\sqrt{a}}{z_{\text{eff}}}^{4/3}(1-b\zeta)^{1/3}.
\end{align}

Since $\zeta<0$, the sign is negative. With the substitution
$\breve{\Lambda}\equiv \left(\frac{\kappa
g\sqrt{{C^2_{T}}}}{{u_\star}^2T\sqrt{a}}\right)^{3/4}$, this leads to
\begin{align}
  &
  z_{\text{eff}}=\frac{1}{{\breve{\Lambda}}}\frac{(-\zeta)^{3/4}}{(1-b\zeta)^{1/4}}
  \rightarrow
  \frac{\zeta}{z_{\text{eff}}}=-{\breve{\Lambda}}(b\zeta^2-\zeta)^{1/4}.\label{zeffunstable2}
\end{align}

We substitute Eq.~(\ref{zeffunstable2}) into Eq.~(\ref{zeffunstable}) to
obtain
\begin{align}
&\zeta =\nonumber\\
&\hack{\hbox\bgroup\fontsize{6.7}{6.7}\selectfont$\displaystyle}\frac{1}{2b}\left(1-\sqrt{
\begin{array}{l}
1+4b{\breve{\Lambda}}(b\zeta^2-\zeta)^{1/4}\\
\cdot\left[\int\limits_0^1{(z(u)+bz(u)^2{\breve{\Lambda}}(b\zeta^2-\zeta)^{1/4})^{-2/3}G(u)\mathrm{d}u}\right]^{-3/2}
\end{array}
}\right) \equiv \breve{F}(\zeta).\hack{$\egroup}\label{zetaunstable}
\end{align}

%f1

This single equation is in the single unknown $\zeta$, since $z(u)$, $G(u)$
and ${\breve{\Lambda}}$ are known; it is also in the fixed-point form
$\zeta=\breve{F}(\zeta)$. The tree diagram for the unstable case is seen in
Fig.~\ref{TREEUNSTABLE}. Evaluation of global partial derivatives proceeds
analogously to the stable case as in Eq.~(\ref{exampleerrorzeta}). Now we
have
\begin{align}
\left(\frac{\delta H_\mathrm{S}}{\delta z (u)}\right) =& \left(\frac{\partial H_\mathrm{S}}{\partial T_\star}\right)
\left(\left(\frac{\partial T_\star}{\partial z_{\text{eff}}}\right)\left(\frac{\partial z_{\text{eff}}}{\partial \zeta}\right)\right.\nonumber\\&\left.+
\left(\frac{\partial T_\star}{\partial \zeta}\right)_{z_{\text{eff}}}
\right)\left(\frac{\delta \zeta}{\delta z (u)}\right).
\end{align}

To pursue the solution of $S_{H_\mathrm{S},z}(u)$, we will need to solve for
$\left(\frac{\partial z_{\text{eff}}}{\partial \zeta}\right)$ by the
differentiation of Eq.~(\ref{zeffunstable2}):
\begin{align}
 &
\left(\frac{\partial z_{\text{eff}}}{\partial \zeta}\right) =\frac{(2b\zeta-3)}{4{\breve{\Lambda}}(-\zeta)^{1/4}(1-b\zeta)^{5/4}} = \frac{z_{\text{eff}}(3-2b\zeta)}{4\zeta(1-b\zeta)}.
\end{align}

We can solve for $\left(\frac{\delta \zeta}{\delta z(u)}\right)$ by implicit
differentiation of Eq.~(\ref{zetaunstable}). In finding $\left(\frac{\delta
\zeta}{\delta z(u)}\right)$, it is useful to define
\begin{align}
&f({\breve{\Lambda}},\zeta(z(u),\breve{\Lambda}),z(u)) \equiv~
1+4b{\breve{\Lambda}}(b\zeta^2-\zeta)^{1/4}\nonumber \\
&\hack{\hbox\bgroup\fontsize{9.5}{9.5}\selectfont$\displaystyle}\cdot\left[\,\int\limits_0^1{(z(u)+bz(u)^2{\breve{\Lambda}}(b\zeta^2-\zeta)^{1/4})^{-2/3}G(u)\mathrm{d}u}\right]^{-3/2},\hack{$\egroup} \label{fdef}
\end{align}
where, from Eqs.~(\ref{zetaunstable}) and (\ref{fdef}), we have
\begin{align}
 &
\sqrt{f}=(1-2b\zeta) \label{tricky}.
\end{align}

From Eq.~(\ref{fdef}), we have
\begin{align}
 &
\left(\frac{\delta f}{\delta z(u)}\right)=\left(\frac{\partial f}{\partial \zeta}\right)\left(\frac{\delta \zeta}{\delta z(u)}\right)+\left(\frac{\delta f}{\delta z(u)}\right)_\zeta,
\end{align}
such that, by implicitly differentiating Eq.~(\ref{tricky}) and then
substituting, we derive
%\hack{\allowdisplaybreaks}
\begin{align}
&\left(\frac{\delta \zeta}{\delta z(u)}\right) =\frac{-\left(\frac{\delta f}{\delta z(u)}\right)_\zeta}{\left(\frac{\partial f}{\partial \zeta}\right)+4b(1-2b\zeta)}, \nonumber \\
&\hack{\hbox\bgroup\fontsize{7.5}{7.5}\selectfont$\displaystyle}=\frac{-\frac{4\zeta(1-b\zeta)}{(1-2b\zeta)}\left(z(u)+bz(u)^2{\breve{\Lambda}}\left(b\zeta^2-\zeta\right)^{\frac{1}{4}}\right)^{-\frac{5}{3}}
\left(1+2bz(u){\breve{\Lambda}}(b\zeta^2-\zeta)^{\frac{1}{4}}\right)G(u)}
{\left\{\left[\int\limits_0^1{\left(z(u)+bz(u)^2{\breve{\Lambda}}(b\zeta^2-\zeta)^{\frac{1}{4}}\right)^{-\frac{2}{3}}G(u)\mathrm{d}u}\right]\right.}\hack{$\egroup}\nonumber\\
&\hack{\hbox\bgroup\fontsize{7.5}{7.5}\selectfont$\displaystyle}\left.\quad +b{\breve{\Lambda}}(b\zeta^2-\zeta)^{\frac{1}{4}}\left[\int\limits_0^1{\left(z(u)+bz(u)^2{\breve{\Lambda}}(b\zeta^2-\zeta)^{\frac{1}{4}}\right)^{-\frac{5}{3}}z(u)^2G(u)\mathrm{d}u}\right]\right.\hack{$\egroup}\nonumber\\
&\hack{\hbox\bgroup\fontsize{7.5}{7.5}\selectfont$\displaystyle}\left.\quad -\frac{4(b\zeta^2-\zeta)^{\frac{3}{4}}}{{\breve{\Lambda}}}\left[\int\limits_0^1{\left(z(u)+bz(u)^2{\breve{\Lambda}}(b\zeta^2-\zeta)^{\frac{1}{4}}\right)^{-\frac{2}{3}}G(u)\mathrm{d}u}\right]^{\frac{5}{2}}\right\}\hack{$\egroup}
\label{dzetadz}
\end{align}

All the information we need to solve for $S_{H_\mathrm{S},z}(u)$ is now
resolved.

\subsection{Full expression for the sensitivity function $S_{H_\mathrm{S},z}(u)$}

Since we are considering an independent $u_\star$ measurement, we have
$S_{T_\star,
z}(u)=S_{H_\mathrm{S},z}(u)=\frac{z(u)}{T_\star}\left(\frac{\delta
T_\star}{\delta z(u)}\right)$. We obtain
% \begin{empheq}[left={S_{T_\star,z}(u) = \empheqlbrace}]{alignat=2}
% %\[
% %
% &\frac{{z(u)}^{-2/3}G(u)}{\int\limits_0^1{{z(u)}^{-2/3}G(u)\mathrm{d}u}}\left(\frac{1}{3+4c\zeta^{2/3}}\right) & \quad\quad\quad \zeta>0 , \label{STZPOSEQ} \\
% %
% &\frac{\begin{array}{l}
% -{z(u)(z(u)+bz(u)^2{\breve{\Lambda}}(b\zeta^2-\zeta)^\frac{1}{4})^{-\frac{5}{3}}}\\
% {\cdot\,(1+2bz(u){\breve{\Lambda}}(b\zeta^2-\zeta)
% ^{\frac{1}{4}})G(u)}
% \end{array}
% }
% {
%  \begin{array}{l}
% \left[\int\limits_0^1{(z(u)+bz(u)^2{\breve{\Lambda}}(b\zeta^2-\zeta)^{\frac{1}{4}})^{-\frac{2}{3}}G(u)\mathrm{d}u}\right]\\
% +b{\breve{\Lambda}}(b\zeta^2-\zeta)^{\frac{1}{4}}\\
% \cdot\left[\int\limits_0^1{(z(u)+bz(u)^2{\breve{\Lambda}}(b\zeta^2-\zeta)^{\frac{1}{4}})^{-\frac{5}{3}}z(u)^2G(u)\mathrm{d}u}\right]\\
% -\frac{4(b\zeta^2-\zeta)^{\frac{3}{4}}}{{\breve{\Lambda}}}\\
% \cdot\,\left[\int\limits_0^1{(z(u)+bz(u)^2{\breve{\Lambda}}(b\zeta^2-\zeta)^{\frac{1}{4}})^{-\frac{2}{3}}G(u)\mathrm{d}u}\right]^{\frac{5}{2}}\\
% \end{array}
% } & \quad\quad\quad \zeta<0 .\label{STZNEGEQ}
% %\
% \end{empheq}
%
\begin{align}
   S_{T_\star,z}(u) =\label{STZPOSEQ}
\end{align}
\begin{eqnarray}
%\begin{empheq}[left= \empheqlbrace]{alignat=1}
\label{STZNEGEQ}\left\{
\hack{\hbox\bgroup\fontsize{9.0}{9.0}\selectfont$\displaystyle}{\begin{array}{lll}
\frac{{z(u)}^{-2/3}G(u)}{\int\limits_0^1{{z(u)}^{-2/3}G(u)\mathrm{d}u}}\left(\frac{1}{3+4c\zeta^{2/3}}\right) &\\ \quad\quad\quad \zeta>0 , \\
\frac{\begin{array}{l}
-{z(u)(z(u)+bz(u)^2{\breve{\Lambda}}(b\zeta^2-\zeta)^\frac{1}{4})^{-\frac{5}{3}}}\\
{\cdot\,(1+2bz(u){\breve{\Lambda}}(b\zeta^2-\zeta)
^{\frac{1}{4}})G(u)}
\end{array}
}
{
 \begin{array}{l}
\left\{\left[\int\limits_0^1{(z(u)+bz(u)^2{\breve{\Lambda}}(b\zeta^2-\zeta)^{\frac{1}{4}})^{-\frac{2}{3}}G(u)\mathrm{d}u}\right]\right.\\
+b{\breve{\Lambda}}(b\zeta^2-\zeta)^{\frac{1}{4}}\\
\left.\cdot\left[\int\limits_0^1{(z(u)+bz(u)^2{\breve{\Lambda}}(b\zeta^2-\zeta)^{\frac{1}{4}})^{-\frac{5}{3}}z(u)^2G(u)\mathrm{d}u}\right]\right.\\
-\frac{4(b\zeta^2-\zeta)^{\frac{3}{4}}}{{\breve{\Lambda}}}\\
\left.\cdot\,\left[\int\limits_0^1{(z(u)+bz(u)^2{\breve{\Lambda}}(b\zeta^2-\zeta)^{\frac{1}{4}})^{-\frac{2}{3}}G(u)\mathrm{d}u}\right]^{\frac{5}{2}}\right\}\\
\end{array}
} &\\ \quad\quad\quad \zeta<0 .\\
\end{array}}\hack{$\egroup}\right.
%\end{empheq}
\end{eqnarray}

\section{Application of the results for the sensitivity function $S_{H_\mathrm{S},z}(u)$}

\subsection{Imnavait Creek basin field campaign}

%f2
\begin{figure}[t]
\includegraphics[width=85mm]{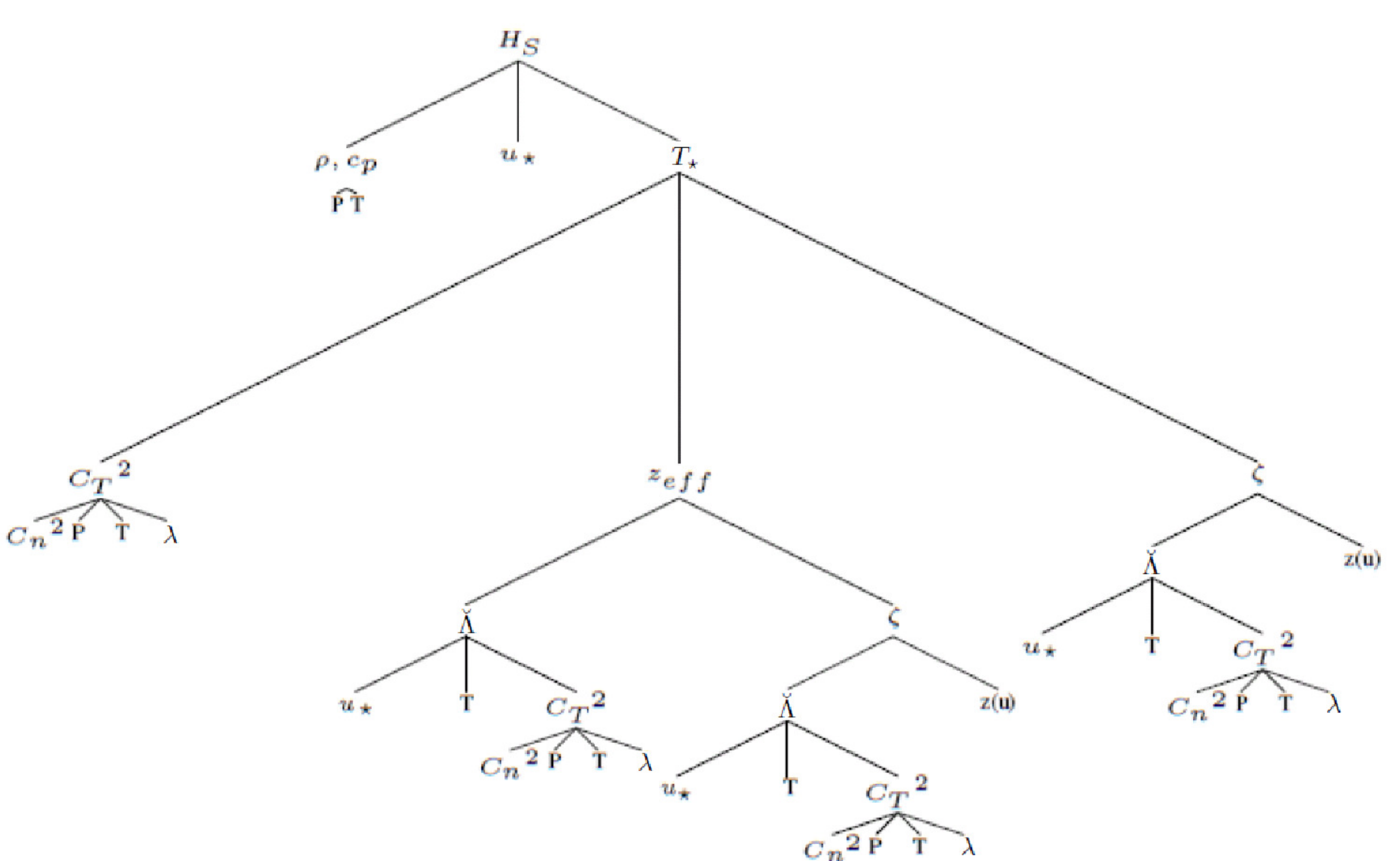}
\caption{Variable inter-dependency tree diagram for the unstable case
($\zeta<0$). The measurement variables are at the end of each branch; all
other variables are derived.} \label{TREEUNSTABLE}
\end{figure}

\begin{figure*}[t]
\includegraphics[width=155mm]{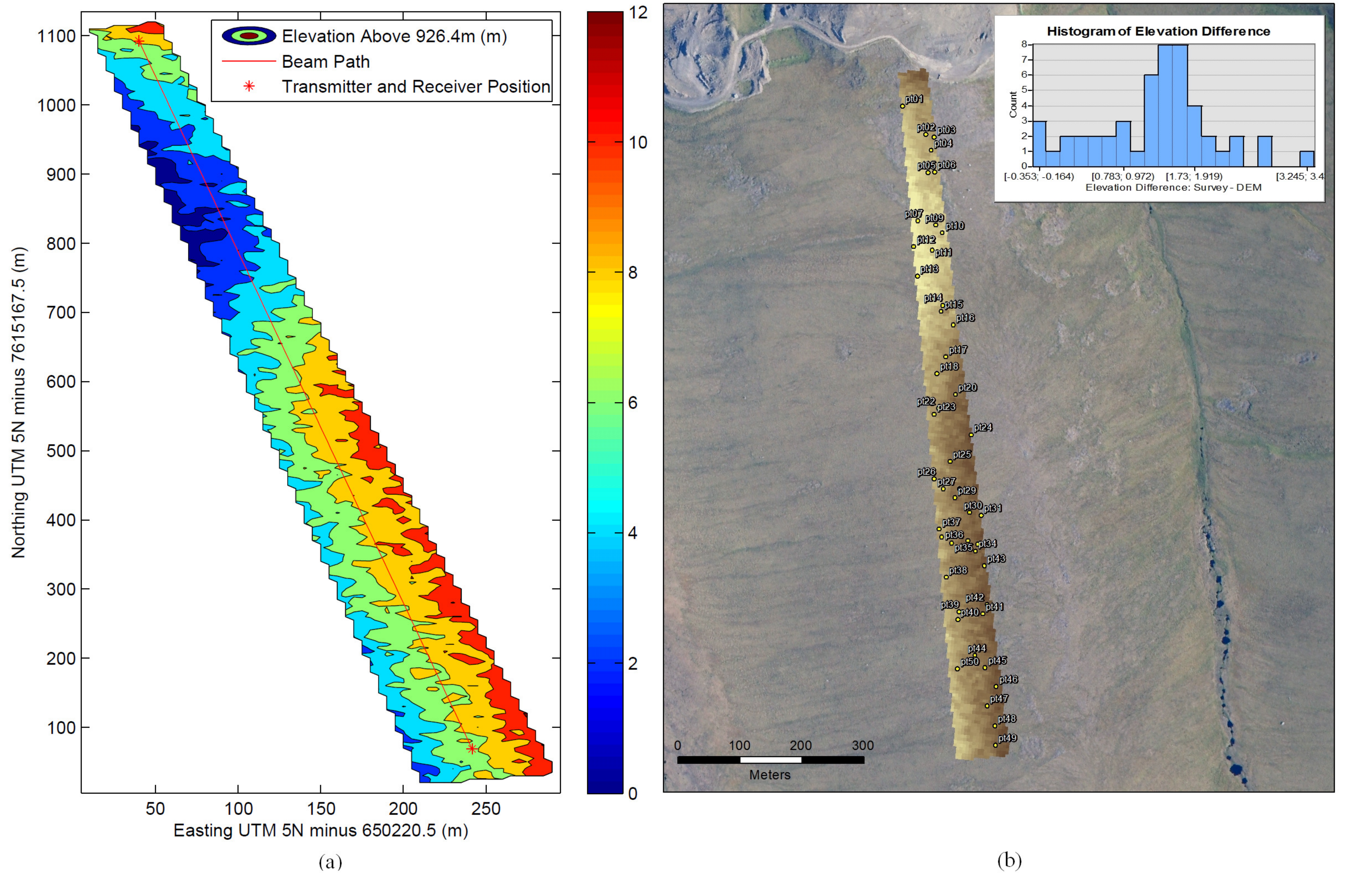}
\caption{Topography and space view of the Imnavait Creek basin, North Slope
of Alaska. The scintillometer beam runs roughly north--south on
a~$1.04$\,\unit{km} path. The emitter and receiver are each raised off the
ground by $3.8$\,\unit{m}. Vegetation along the path is representative of
Arctic tundra. Superimposed is a~histogram of $50$ points of the GPS ground
truth elevation survey minus the DEM elevation.} \label{BIGFIG}
\end{figure*}

\begin{figure}[p]
\includegraphics[width=85mm]{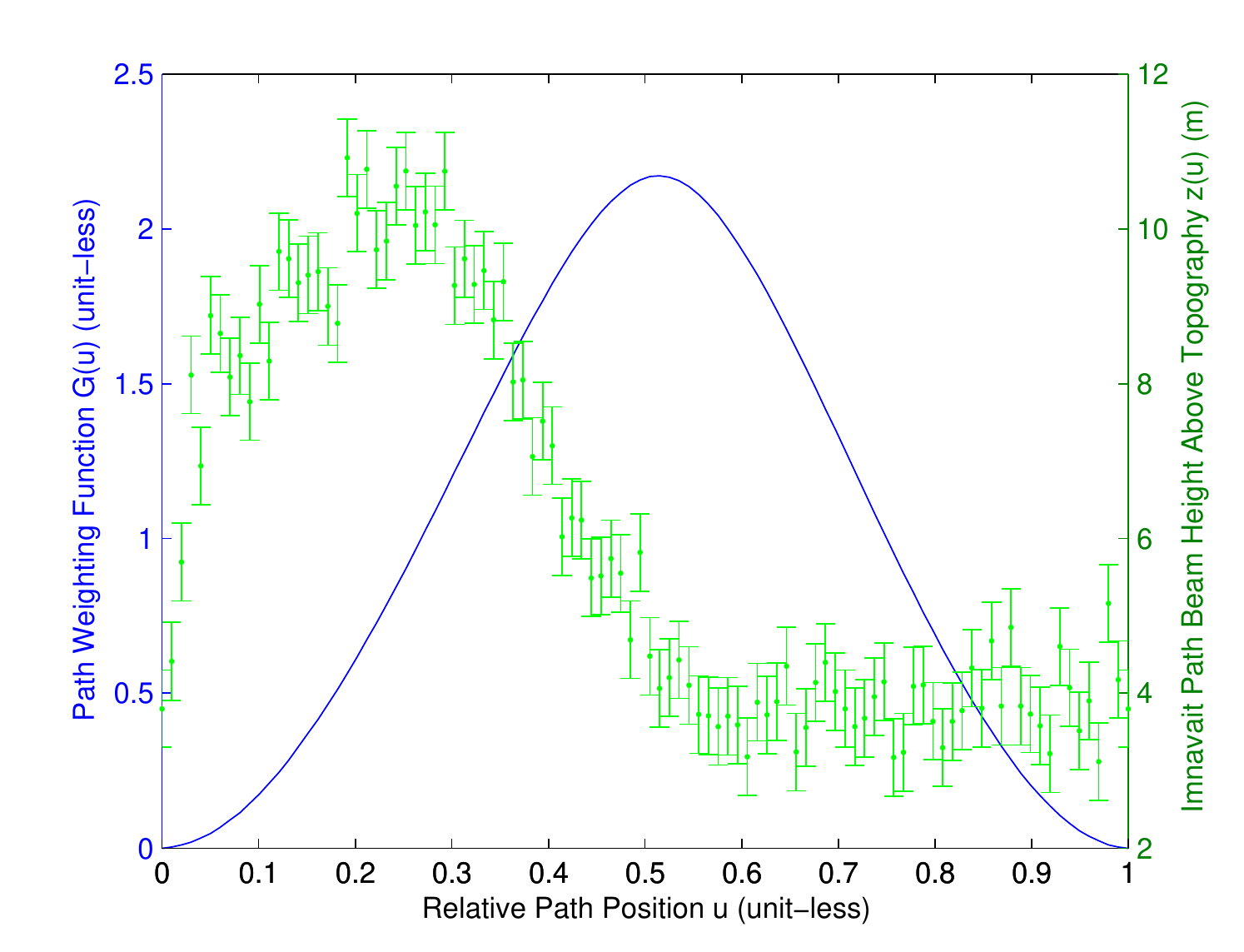}
\caption{Height of the beam above the ground $z$ and the path weighting
function $G$ as functions of relative path position $u$, using the Imnavait
experimental site as seen in Fig.~\ref{BIGFIG}a. Uncertainties are based on
the approximate standard deviation in the histogram in Fig.~\ref{BIGFIG}b,
although they do not influence the analysis presented in this study.}
\label{zG}
\end{figure}

As an example, we use topographic data from the Imnavait Creek basin field
site (UTM 5N 650220.5 East, 7615761.5 North), where there is a~campaign to
determine large-scale turbulent fluxes in the Alaskan tundra; it is seen in
Figs.~\ref{BIGFIG}a and \ref{zG}. We assume for simplicity that vegetation
patterns, water availability, and other changes across the basin that could
affect the flow in the atmospheric surface layer do not represent
a~significant source of surface heterogeneity. The elevation data seen in
Fig.~\ref{BIGFIG}a are from a~$5$\,\unit{m} resolution digital elevation map
(DEM), which has a~roughly $0.5$\,\unit{m} standard deviation in a~histogram
of the difference between the DEM elevations and $50$ randomly distributed
GPS ground truth points, as seen in Fig.~\ref{BIGFIG}b. Note that the
systematic offset between the DEM and the GPS ground truth measurements does
not contribute to systematic error in $z(u)$. Note also that some of this
spread in data may be due to an active permafrost layer.

%f3

\begin{figure}
\includegraphics[width=85mm]{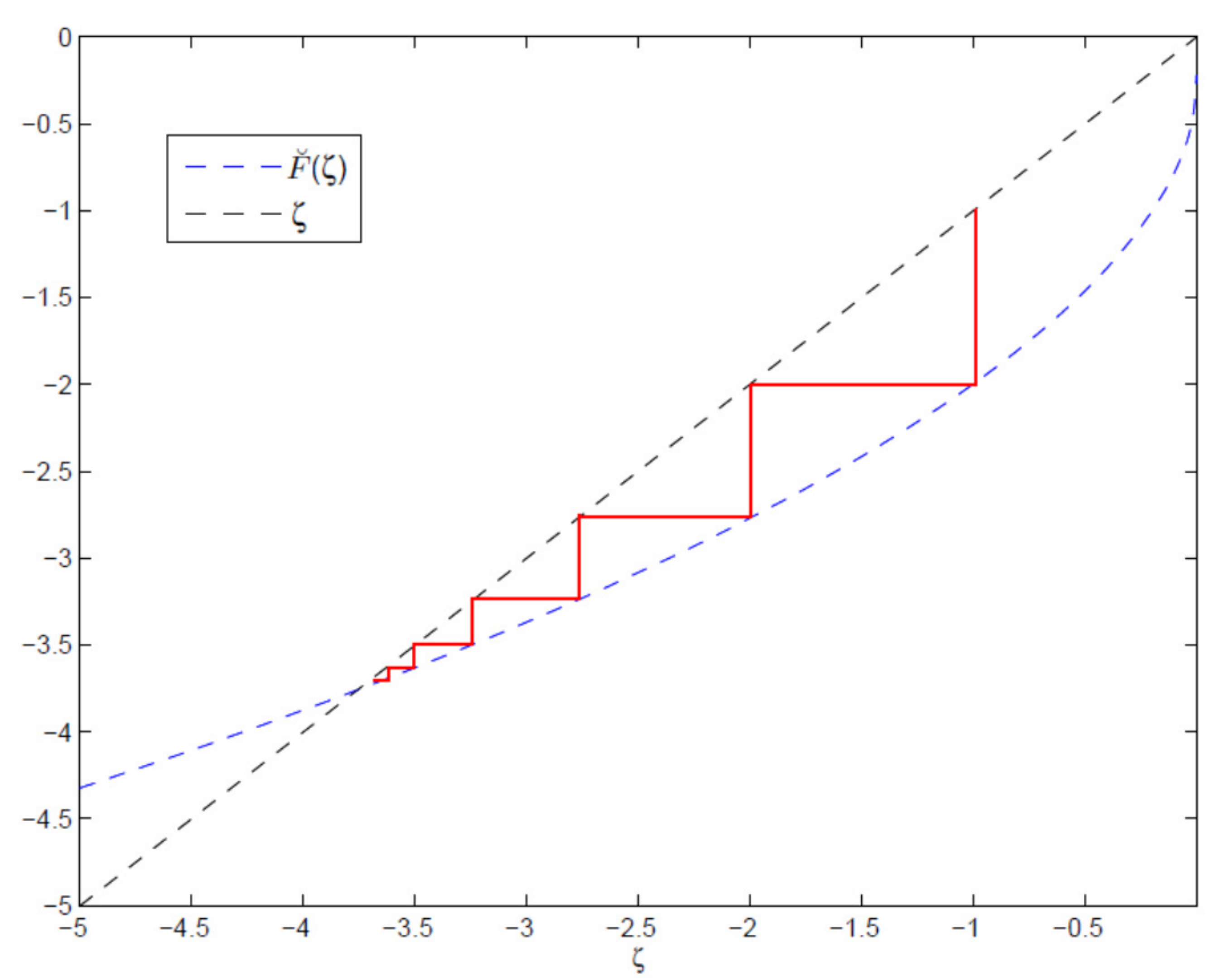}
\caption{Graphical visualization of the fixed-point solution of
Eq.~(\ref{zetaunstable}). The recursively defined series
$[\breve{F}(\zeta_{\text{guess}}),\breve{F}(\breve{F}(\zeta_{\text{guess}})),\breve{F}(\breve{F}(\breve{F}(\zeta_{\text{guess}}))),\dots]$
converges monotonically for any $\zeta_{\text{guess}}<0$. A~typical value of
${\breve{\Lambda}}=1/4$ is used, representing slightly unstable conditions in
the atmospheric surface layer. The initial guess is
$\zeta_{\text{guess}}=-1$, and the path of convergence is shown by the red
line. The Imnavait Creek basin terrain and beam path are used for $z(u)$,
along with the standard path weighting function $G(u)$ as seen in
Figs.~\ref{BIGFIG}a and \ref{zG}.} \label{ZETASOLFIG}
\end{figure}

For this field site, we can solve for $\zeta$ under unstable conditions
through Eq.~(\ref{zetaunstable}). As can be seen in Fig. \ref{ZETASOLFIG}, we
arrive at the solution for $\zeta$ with the recursively defined series
$[\breve{F}(\zeta_{\text{guess}}),\breve{F}(\breve{F}(\zeta_{\text{guess}})),\breve{F}(\breve{F}(\breve{F}(\zeta_{\text{guess}}))),\dots]$
that is guaranteed to converge monotonically for any
$\zeta_{\text{guess}}<0$.

\begin{figure}[t]
\includegraphics[width=85mm]{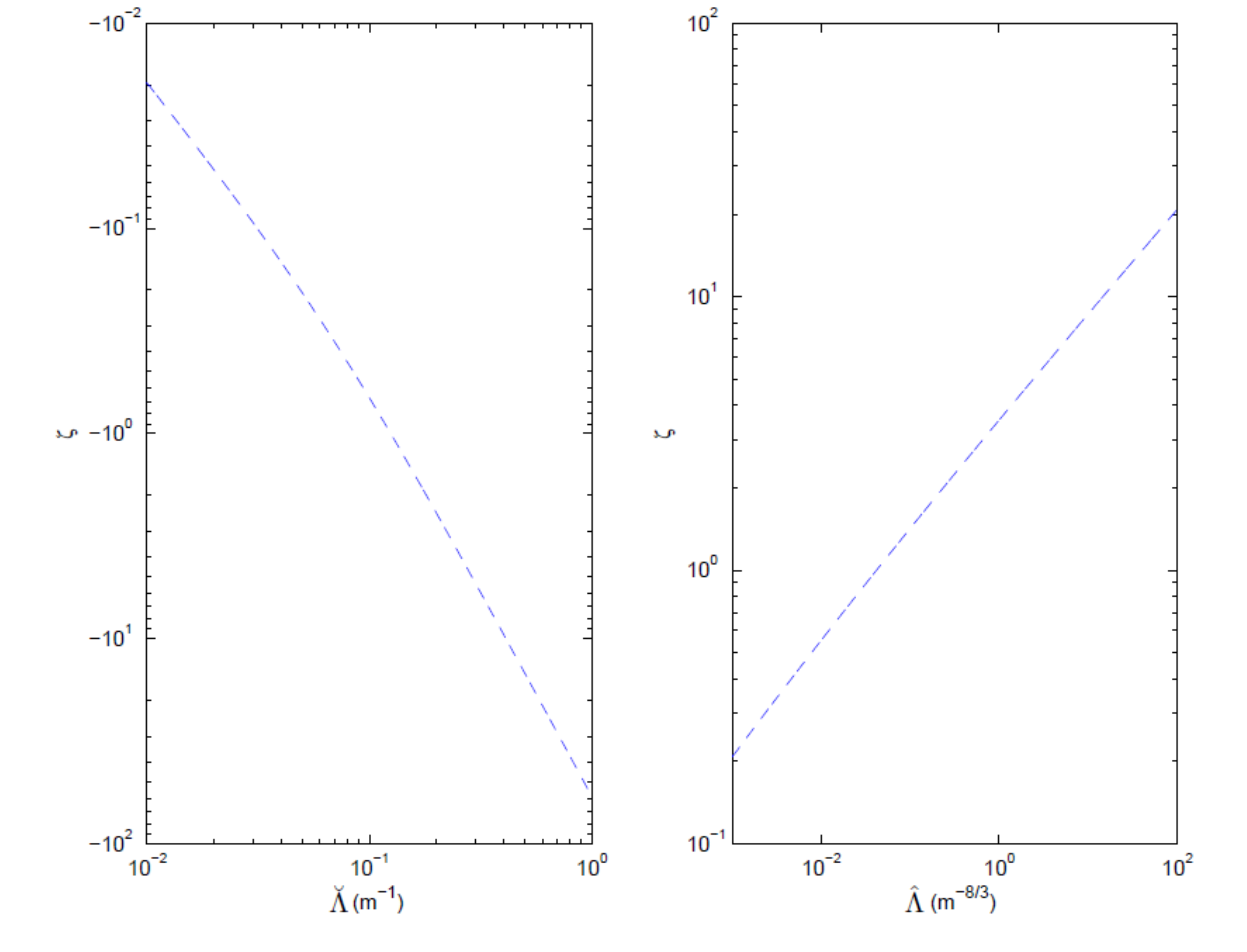}
\caption{Solution of Eqs.~(\ref{stablerecursion}) and (\ref{zetaunstable})
produced with a~monotonically converging series as explained in the text and
as visualized in Fig. \ref{ZETASOLFIG}. The Imnavait Creek basin terrain and
beam path are used for $z(u)$, along with the standard path weighting
function $G(u)$ as seen in Figs.~\ref{BIGFIG}a and \ref{zG}. The mapping
between $\zeta$ and $\breve{\Lambda}$ and between $\zeta$ and $\hat{\Lambda}$
is bijective. Note that the solution of $\zeta$ for $\breve{\Lambda}=1/4$
corresponds to the intersection of $\breve{F}$ with $\zeta$ in
Fig.~\ref{ZETASOLFIG}.} \label{PHIVSZETA}
\end{figure}

A~plot of $\zeta$ as a~function of ${\breve{\Lambda}}$ for this field site is
seen in Fig.~\ref{PHIVSZETA}. Note that the relationship between $\zeta$ and
${\breve{\Lambda}}$ is bijective; any value of ${\breve{\Lambda}}$ is
uniquely associated with a~value of $\zeta$.

%f4

%f5

%f6

%f7
\begin{figure}[t]
\includegraphics[width=85mm]{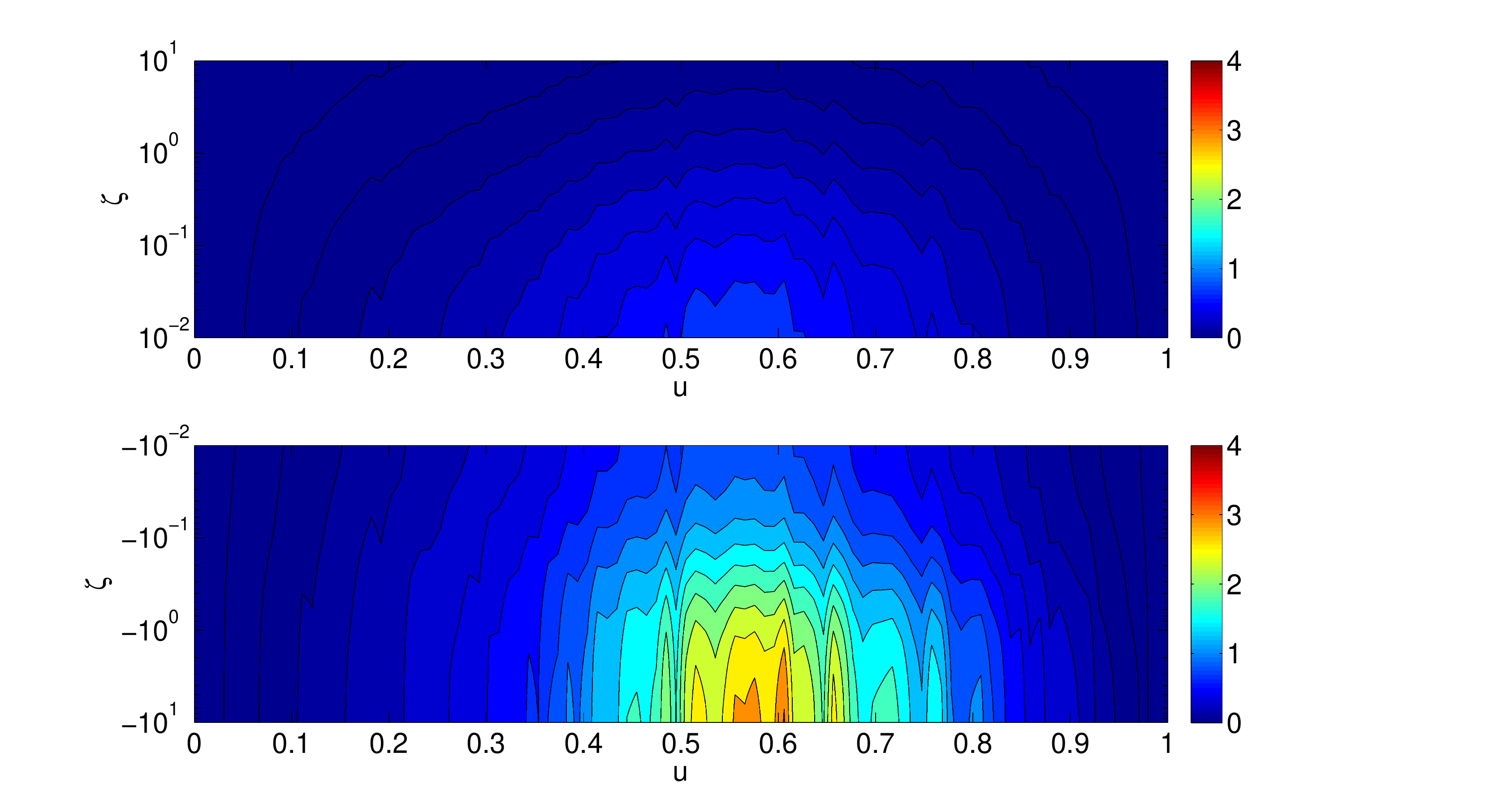}
\caption{Sensitivity function $S_{H_\mathrm{S},z}(u)=S_{T_\star, z}(u)$. For
stable conditions ($\zeta>0$), $S_{T_\star, z}(u)$ is given in
Eq.~(\ref{STZPOSEQ}). For unstable conditions ($\zeta<0$), $S_{T_\star,
z}(u)$ is given by Eq.~(\ref{STZNEGEQ}), where values for $\zeta$ as
a~function of ${\breve{\Lambda}}$ are obtained through a~numerical solution
of Eq.~(\ref{zetaunstable}), which may be visualized with Fig.
\ref{PHIVSZETA}. The Imnavait Creek basin terrain and beam path are used for
$z(u)$, along with the standard path weighting function $G(u)$ as seen in
Figs.~\ref{BIGFIG}a and \ref{zG}.} \label{STZK}
\end{figure}

Considering the field case study of the Imnavait Creek basin, where the
height of the beam over the terrain $z(u)$ and the standard path weighting
function $G(u)$ are seen in Figs.~\ref{BIGFIG}a and \ref{zG},
Eqs.~(\ref{STZPOSEQ}) and (\ref{STZNEGEQ}) lead to the sensitivity function
seen in Fig.~\ref{STZK}. Note that $S_{H_\mathrm{S},z}(u)$ is a~function of
$u$ and $\zeta$ only, since, for any one beam height transect $z(u)$,
$\breve{\Lambda}$ is mapped bijectively with respect to $\zeta$ through
Eq.~(\ref{zetaunstable}), as seen in Fig.~\ref{PHIVSZETA}.

Note that if we consider a~constant ratio of $\frac{\sigma_z(u)}{z(u)}$,
systematic error propagation can be re-written as
\begin{align}
 &
\int\limits_0^1{\frac{\sigma_z(u)}{z(u)}S_{H_\mathrm{S},z}(u)\mathrm{d}u}
= \frac{\sigma_z(u)}{z(u)}\left[\,\int\limits_0^1{S_{H_\mathrm{S},z}(u)\mathrm{d}u}\right].\label{CONSTRATIO}
\end{align}

The term in square brackets on the right of Eq.~(\ref{CONSTRATIO}) is plotted
in Fig.~\ref{STZCONSTRATIO}.

\subsection{Synthetic scintillometer beam paths}

It is interesting to examine the sensitivity function over synthetic paths
that are representative of commonly used paths in scintillometry. Two
synthetic paths can be seen in Fig.~\ref{SyntheticPaths}. They include
a~slant path as well as a~quadratic path representing a~beam over a~valley.

The sensitivity function $S_{T_\star, z}(u)=S_{H_\mathrm{S},z}(u)$ for
synthetic path $1$ (the quadratic path) seen in Fig.~\ref{SyntheticPaths} is
seen in Fig.~\ref{quadratic}. For synthetic path $2$ (the slant path), the
sensitivity function is seen in Fig.~\ref{linear}.

\hack{\newpage}

\section{Discussion}

A~sensitivity function mapping the propagation of uncertainty from $z(u)$ to
$H_\mathrm{S}$ has been produced for a~large-aperture scintillometer strategy
incorporating independent $u_\star$ measurements, and the line integral
footprint approach to variable topography developed in
\citet{HARTOGENSIS2003} and \mbox{\citet{KLEISSL2008}}. This was accomplished
by mapping out the variable inter-dependency as illustrated in the tree
diagrams in Figs.~\ref{TREESTABLE} and \ref{TREEUNSTABLE}, and by applying
functional derivatives. The solution to $S_{H_\mathrm{S},z}(u)$ is given in
Eqs.~(\ref{SHZ}), (\ref{STZPOSEQ}) and (\ref{STZNEGEQ}).

\begin{figure}[t]
\includegraphics[width=85mm]{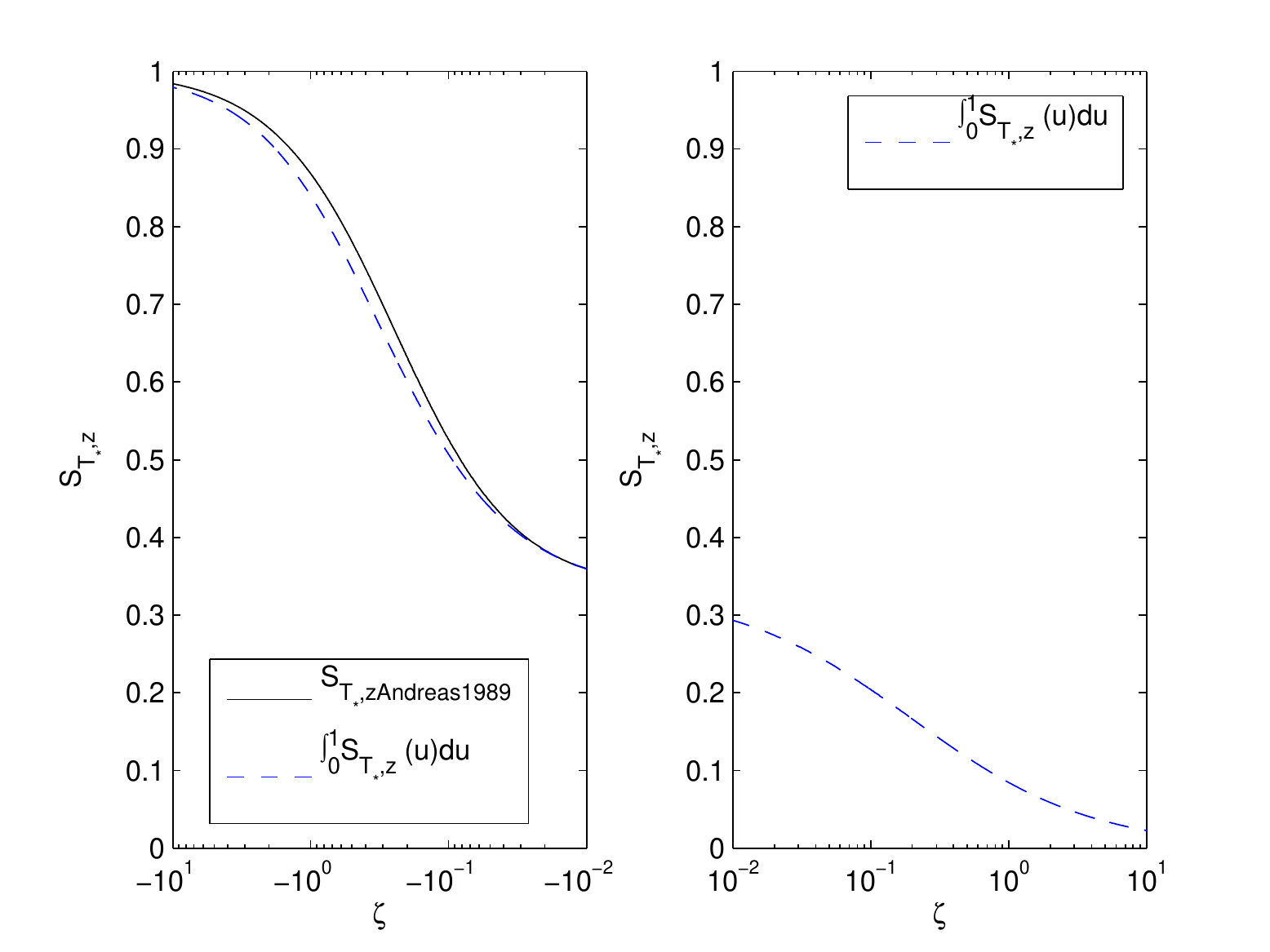}
\caption{Average value of $S_{T_\star, z}(u)=S_{H_\mathrm{S},z}(u)$ over beam
path $u$, given by $\int_0^1{S_{T_\star, z}(u)\mathrm{d}u}$, and the flat
terrain sensitivity function $S_z$ derived in \citet{ANDREAS1989} (for
$\zeta>0$, the functions are identical). For stable conditions ($\zeta>0$),
$S_{T_\star, z}(u)$ is given by Eq.~(\ref{STZPOSEQ}). For unstable conditions
($\zeta<0$), $S_{T_\star, z}(u)$ is given by Eq.~(\ref{STZNEGEQ}), where
values for $\zeta$ as a~function of ${\breve{\Lambda}}$ are obtained through
a~numerical solution of Eq.~(\ref{zetaunstable}), which may be visualized in
Fig. \ref{PHIVSZETA}. The Imnavait Creek basin terrain and beam path are used
for $z(u)$, along with the standard path weighting function $G(u)$ as seen in
Figs. \ref{BIGFIG}a and \ref{zG}.} \label{STZCONSTRATIO}
\end{figure}

\begin{figure}[t]
\includegraphics[width=85mm]{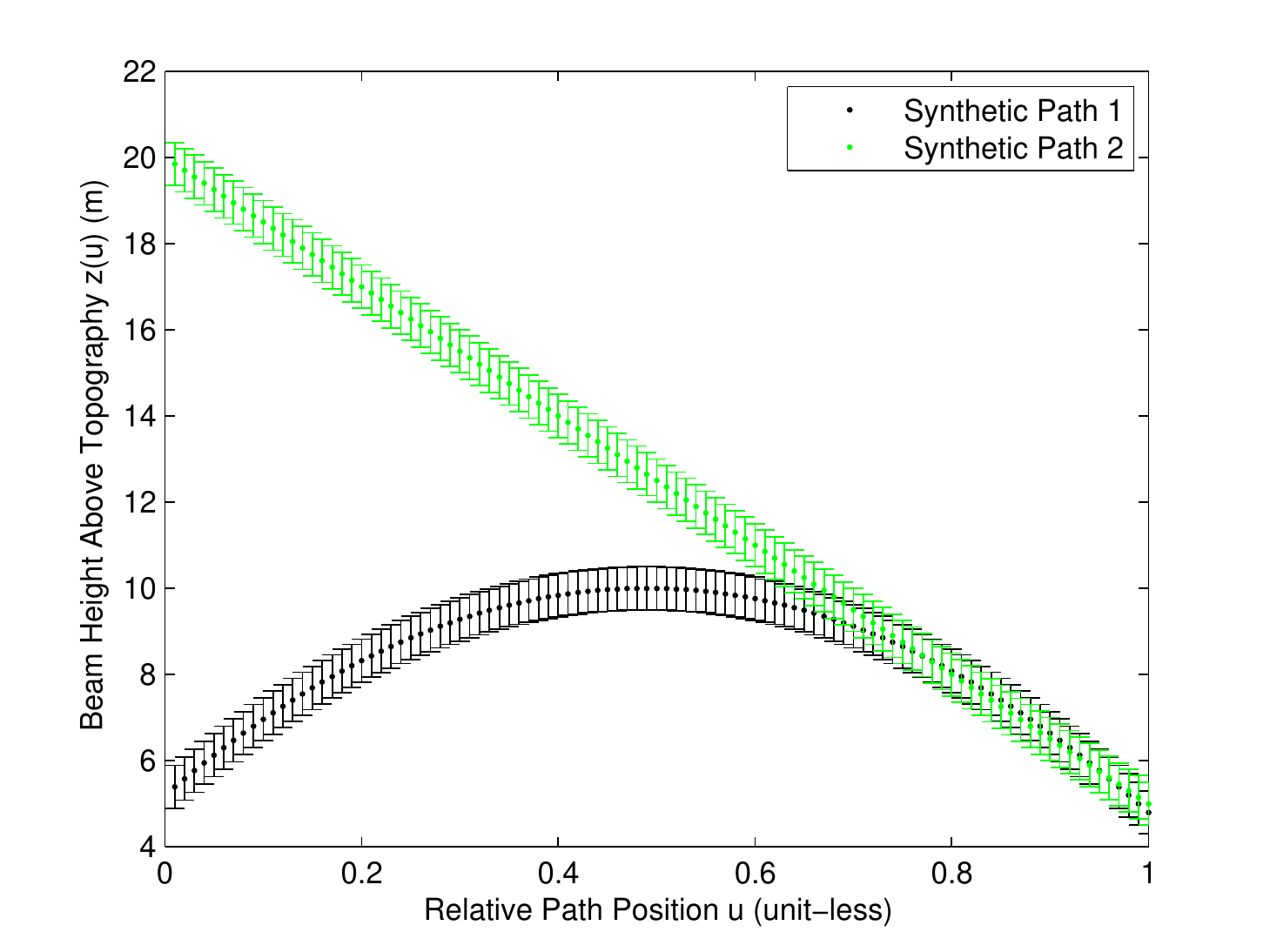}
\caption{Synthetic path beam heights including a~quadratic path (path 1) and
a~slant path (path 2).} \label{SyntheticPaths}
\end{figure}

\begin{figure}[t]
\includegraphics[width=85mm]{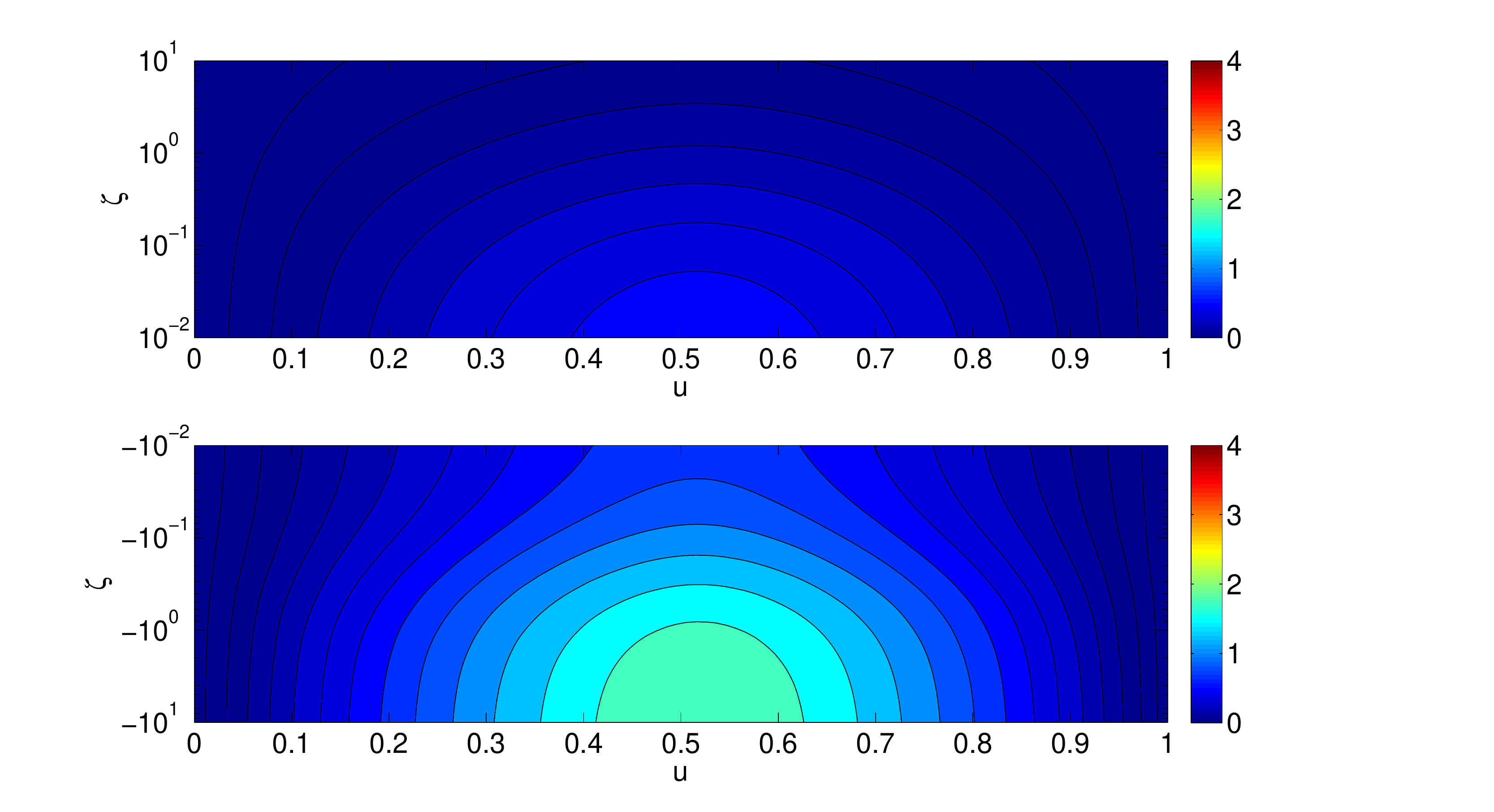}
\caption{Sensitivity function $S_{H_\mathrm{S},z}(u)=S_{T_\star, z}(u)$. For
stable conditions ($\zeta>0$), $S_{T_\star, z}(u)$ is given in
Eq.~(\ref{STZPOSEQ}). For unstable conditions ($\zeta<0$), $S_{T_\star,
z}(u)$ is given by Eq.~(\ref{STZNEGEQ}), where values for $\zeta$ as
a~function of ${\breve{\Lambda}}$ are obtained through a~numerical solution
of Eq.~(\ref{zetaunstable}), which may be visualized with
Fig.~\ref{PHIVSZETA}. Synthetic beam path $1$ (the quadratic path) is used
for $z(u)$, along with the standard path weighting function $G(u)$ as seen in
Figs.~\ref{SyntheticPaths} and \ref{zG}.} \label{quadratic}
\end{figure}

\begin{figure}
\includegraphics[width=85mm]{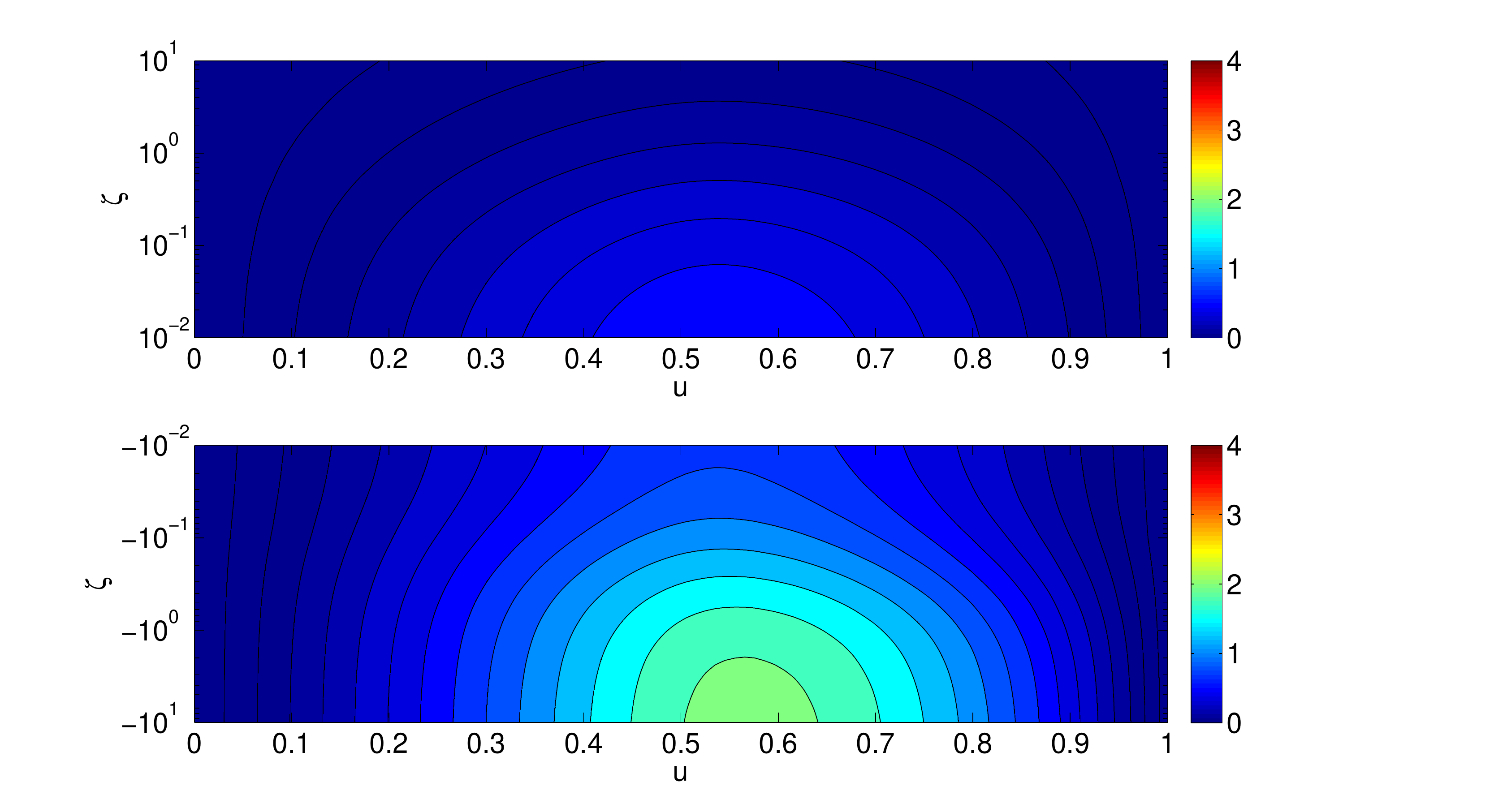}
\caption{Sensitivity function $S_{H_\mathrm{S},z}(u)=S_{T_\star, z}(u)$. For
stable conditions ($\zeta>0$), $S_{T_\star, z}(u)$ is given in
Eq.~(\ref{STZPOSEQ}). For unstable conditions ($\zeta<0$), $S_{T_\star,
z}(u)$ is given by Eq.~(\ref{STZNEGEQ}), where values for $\zeta$ as
a~function of ${\breve{\Lambda}}$ are obtained through a~numerical solution
of Eq.~(\ref{zetaunstable}), which may be visualized with Fig.
\ref{PHIVSZETA}. Synthetic beam path $2$ (the slant path) is used for $z(u)$,
along with the standard path weighting function $G(u)$ as seen in Figs.
\ref{SyntheticPaths} and \ref{zG}.} \label{linear}
\end{figure}

As seen in Figs.~\ref{BIGFIG}a, \ref{zG}, and \ref{STZK}, our results for
$S_{T_\star, z}(u)=S_{H_\mathrm{S},z}(u)$ show that sensitivity to
uncertainties in topographic heights is generally higher
under  unstable conditions, and it is both concentrated in the center of the
path and in areas where the underlying topography approaches the beam height.
This finding intuitively makes sense, since scintillometers are more
sensitive to $C_{T}^2$ at the center of their beam path, and $C_{T}^2$
decreases nonlinearly in height above the surface and strengthens with
greater instability. For the Imnavait Creek basin path, the value of
$S_{H_\mathrm{S},z}(u)$ increases to $3$ at small dips in the beam height
beyond the halfway point of the path, as seen in Fig.\ref{STZK}. Note that
the asymmetry along $u$ of $S_{H_\mathrm{S},z}(u)$ corresponds to the
asymmetry of the path, which is mostly at a~higher ($>$\,6\,\unit{m}) height
in the first half, and at a~lower height ($\approx$\,4\,\unit{m}) in the
second half, as seen in Fig.~\ref{zG}. Also note that the local maxima in
$S_{H_\mathrm{S},z}(u)$ occur at roughly $u\approx$\,60\,{\%} and
$u\approx$\,65\,{\%}; these correspond directly to topographic protuberances
seen in Figs.~\ref{BIGFIG}a and \ref{zG}. Note that the total error in
$H_\mathrm{S}$ is contributed from the whole range of $u$ along
$S_{H_\mathrm{S},z}(u)$, so even though we may have values of up to $3$ in
the sensitivity functions, our error bars may still be reasonable. The
average value of $S_{H_\mathrm{S},z}(u)$ along $u$ is never higher than $1$,
as seen in Fig. \ref{STZCONSTRATIO}. Knowledge of where the concentration in
sensitivity is allows us to decrease our uncertainty greatly by taking
high-accuracy topographic measurements in these areas, especially for Arctic
beam paths, which must be low due to thin boundary layers.

%f8

For example, if the random error in $z(u)$ in the Imnavait Creek basin were
0.5\,m, the relative error resulting in $H_\mathrm{S}$ due to uncertainty in
$z(u)$ alone would be just 2\,\% under slightly unstable conditions where
$\breve{\Lambda}=1/4$ and $\zeta \approx -3.75$, whereas if we reduce the
uncertainty in  $z(u)$ to
0.1\,m, the relative error in $H_\mathrm{S}$ due to uncertainty in $z(u)$
would be just 0.3\,\%, so
with a reasonable number of survey points (100), the error can be quite
small. However, if we look at Fig.~\ref{BIGFIG}b, we see that there is
significant systematic error, perhaps due to shifting permafrost. If we have
a perfectly even systematic error across the whole map, then this error is
not propagated. However, if we have even a small amount of systematic error
such as 0.5\,m distributed around the center of the beam path near the local
maxima in sensitivity, we can easily achieve 10\,\% to 20\,\% relative error
in $H_\mathrm{S}$. In comparison to other variables, the values for
$S_{H_\mathrm{S},u_\star}$ are similar in magnitude to $S_{H_\mathrm{S},z}$
under unstable conditions, smaller under neutral conditions, and larger under
stable conditions \citep{ANDREAS1989}. Under unstable conditions, error from
$u_\star$ may therefore be similar in magnitude to error from $z(u)$;
however, for \mbox{path-averaged} $u_\star$ scintillometer strategies, this is not
an issue. For $C_n^2$, the sensitivity functions are usually smaller, but in
isolated regions they are larger \citep{ANDREAS1989}.
%f9

The average value of $S_{H_\mathrm{S},z}(u)$ over the beam path reduces to
identical results to the flat terrain sensitivity function $S_z$ from
\citet{ANDREAS1989} (which would be denoted $S_{T_\star, z}$ here) under
stable conditions where $z_{\text{eff}}$ is de-coupled from $\zeta$, and
nearly identical results (depending on the path) under unstable conditions
where $z_{\text{eff}}$ is coupled to $\zeta$, as seen in
Fig.~\ref{STZCONSTRATIO}. It is unknown as to whether the addition of
equations for path-averaged $u_\star$ measurements such as the Businger--Dyer
relation seen in \citet{HARTOGENSIS2003} and \citet{SOLIGNAC}, or
displaced-beam scintillometer strategies as seen in \citet{ANDREAS1992},
would change these results significantly.

We note that the study of \citet{HARTOGENSIS2003} evaluated a~function
similar to $S_{H_\mathrm{S},z}$ for flat terrain with an independent
$u_\star$ measurement (the 2003 Eq.~7 is ignored); however, at $\zeta\approx
0$ they found a~sensitivity of $1/2$ instead of $1/3$ as found in
\citet{ANDREAS1989}. The difference in the results between these two studies
is not due to the differences between single- and double-wavelength
strategies. The Obukhov length (denoted by $L_{\text{MO}}$ in
\citealp{HARTOGENSIS2003}) is a~function of $z_{\text{LAS}}$ through the 2003
Eqs.~(5) and (6). The addition of chain rule terms to reflect the dependence
of $l$ on $z$ in Hartogensis et al.'s (2003) Eq.~(A2) resolves differences
between Hartogensis et al.'s (2003) Fig.~A1 and Andreas et al.'s (1989)
Fig.~4; the flat-terrain sensitivity function for $\zeta<0$ is
 \begin{align}
  & S_{H_\mathrm{S},z}=S_{T_\star,
    z}=\frac{1-2b\zeta}{3-2b\zeta}\neq\frac{1-2b\zeta}{2-2b\zeta}=\frac{z}{H_\mathrm{S}}\left(\frac{\partial
      H_\mathrm{S}}{\partial z}\right)_l,
\end{align}
which is given correctly in \citet{ANDREAS1989}.

%f10

\hack{\newpage}

Equations~(\ref{errorpropch3}), (\ref{SENSITIVITYEQch3}), (\ref{STZPOSEQ}),
and (\ref{STZNEGEQ}) may be implemented into computer code for routine
analysis of data. It is worth noting that the sign of $\zeta$ is an a priori
unknown from the measurements. Thus, for any set of measurements, we should
calculate the set of all derived variables and their respective uncertainties
assuming both stable and unstable conditions, and if uncertainties in the
range of $\zeta$ overlap with $\zeta=0$ for either stability regime, we
should then consider the combined range of errors in the two sets.

In the application of Eq.~(\ref{errorpropch3}), we must recognize
computational error $\sigma_{f_\mathrm{c}}$. Previous studies have
incorporated a~cyclically iterative algorithm that may not converge, as seen
in \citet{ANDREAS2012}, or that may converge to an incorrect solution, as
illustrated in the section on coupled nonlinear equations in
\citet{PRESSNUM}. We have developed techniques to eliminate this error. For
unstable cases ($\zeta<0$), the solution of $\zeta$ follows from
Eq.~(\ref{zetaunstable}), which is in fixed-point form. The solution to
Eq.~(\ref{zetaunstable}) is guaranteed to converge monotonically with the
recursively defined series
$[\breve{F}(\zeta_{\text{guess}}),\breve{F}(\breve{F}(\zeta_{\text{guess}})),\breve{F}(\breve{F}(\breve{F}(\zeta_{\text{guess}}))),\dots]$
as seen in \citet{ITERATIVE} and in \citet{FIXEDPOINT}, and as demonstrated
in Fig. \ref{ZETASOLFIG}. We may solve for the stable case ($\zeta>0$)
recursively using Eq.~(\ref{stablerecursion}), where $\hat{F}(\zeta)$
demonstrates convergence properties that are similar to those of
$\breve{F}(\zeta)$ in Eq.~(\ref{zetaunstable}). It was found to be practical
to make $\zeta_{\text{guess}} = \pm1$.

%f11

Future expansions of the results presented here should focus on including
multiple wavelength strategies to evaluate the latent heat flux and
$H_\mathrm{S}$, as well as on including path-averaged $u_\star$ measurements
using $l_{o}$ and $C_n^2$ scintillometer strategies as in \citet{ANDREAS1992}
or using a~point measurement of wind speed and the roughness length via the
Businger--Dyer relation \citep[e.g.,][]{PANOFSKY,SOLIGNAC}. Modification of
the analysis for including path-averaged $u_\star$ measurements involves the
addition of one or two more equations \citep[e.g., Eq.~8 in][or Eqs.~1.2 and
1.3]{SOLIGNAC} in \citealp{ANDREAS1992}) to substitute into
Eqs.~(\ref{zetastableeq}) and (\ref{zetaunstable}), as well as the definition
of new tree diagrams to reflect that $u_\star$ is now a~derived variable. In
these cases, either the turbulence inner-scale length $l_{o}$ or a~point
measurement of wind speed and the roughness length replaces
$u_\star$ as a measurement; $u_\star$ is derived through information from the
full set of measurements. Note that if $u_\star$ is derived through
measurements including $z$, Eq.~(\ref{SENSIBLEHEATch3}) implies that
$S_{H_\mathrm{S},z}=S_{T_\star, z}+S_{u_\star, z}$. It is worth investigating
whether computational error can still be eliminated in these cases.

We have considered here the effective height line integral approach derived
in \citet{HARTOGENSIS2003} and \citet{KLEISSL2008} to take into account
variable topography. Even if we assume a~constant flux surface layer, under
realistic wind conditions, turbulent air is advected in from nearby
topography. For example, in the Imnavait Creek basin path seen in
Fig.~\ref{BIGFIG}a, if wind comes from the west, the turbulent air being
advected into the beam path comes from a~volume that is higher above the
underlying topography than if wind came from the east. Sensitivity studies
should be produced for two-dimensional surface integral methods that take
into account the coupling of wind direction and topography on an instrument
footprint \citep[e.g.,][]{MEIJNINGERPWF2002,LIU}. Additionally, a new theory
may be developed for heterogeneous terrain involving complex distributions of
water availability and roughness length such as the terrain in the Imnavait
Creek basin.

\hack{\newpage}
\conclusions

Sensitivity of the sensible heat flux measured by scintillometers has been
shown to be highly concentrated in areas near the center of the beam path and
in areas of topographic protrusion. The general analytic sensitivity
functions that have been evaluated here can be applied for error analysis
over any field site as an alternative to complicated numerical methods.
Uncertainty can be greatly reduced by focusing accurate topographic
measurements in areas of protrusion near the center of the beam path. The
magnitude of the uncertainty is such that it may be necessary to use
high-precision LIght Detection And Ranging (LIDAR)
topographic data as in \citet{GELI2012} for Arctic field sites in order to
avoid large errors resulting from uneven permafrost changes since the last
available DEM was taken. Additionally, computational error can be eliminated
by following a~computational procedure as outlined here.

\begin{acknowledgements}
Matthew Gruber thanks the Geophysical Institute at the University of Alaska
Fairbanks for its support during his Masters degree. We thank Flora Grabowska
of the Mather library for her determination in securing funding for open
access fees, Jason Stuckey and Randy Fulweber at ToolikGIS, Chad Diesinger at
Toolik Research Station, and Matt Nolan at the Institute for Northern
Engineering for the digital elevation map of Imnavait, GPS ground truth
measurements, and Fig.~\ref{BIGFIG}b. G.~J.~Fochesatto was partially
supported by the Alaska Space Grant NASA-EPSCoR program award number
NNX10N02A.\hack{\newline}\hack{\newline}Edited by: M.~Nicolls
\end{acknowledgements}

\end{document}